Warren Leffler,
Department of Mathematics,
Los Medanos College,
2700 East Leland Road, Pittsburg,
CA 94565
wleffler@losmedanos.edu

# Locality And The Path Integral

*ABSTRACT*

We analyze the property of locality with respect to the framework for quantum mechanics based on the path integral formalism. As is well known, this framework makes the same experimental predictions as does the one based on a separable Hilbert space and the Schrödinger equation.

## I. AN ARTIFACT OF A PARTICULAR FORMALISM

Here is a short statement of the general content of Bell's theorem [1]:

> [Bell] reasoned that if any manifestly and completely local algorithm existed that made the same predictions for the outcomes of experiments as the quantum-mechanical algorithm does, then Einstein … would have been right to dismiss the nonlocalities in quantum mechanics as merely an artifact of that particular formalism. Conversely, if no algorithm could avoid nonlocalities, then they must be genuine physical phenomena. Bell then analyzed a specific entanglement scenario and concluded that no such local algorithm was mathematically possible.

His impossibility proof is a derivation of certain simple inequalities—"Bell's inequality" and the CHSH inequality—that are to hold in any local framework for physics, but which fail to hold in quantum mechanics. We will analyze this proof from the viewpoint of the Feynman path integral (FPI) algorithm.

As is well known, the FPI algorithm makes the same predictions for the outcomes of experiments as the standard quantum-mechanical algorithm does, which is based on a separable Hilbert space and the Schrödinger equation. The path integral yields "results identical to those obtained by the standard methods of wave mechanics or matrix manipulation [2]." And, as Stephen Weinberg notes, "the path-integral formalism allows us to find the solution of the Schrödinger equation, without ever writing down the Schrödinger equation [3]." We will assume that the reader is already familiar with the standard Hilbert-space approach, but will now present a brief summary of the FPI formalism.

### (a) The Feynman path-integral framework

Let $\Lambda = C_{u,v}^{0,t}$ be the set of possible paths that a particle could take in going from *u* to *v*—that is, $\Lambda$ is the set of possible paths $x(w)$ on $[0, t]$ mapping 0 to *u* and *t* to *v*.

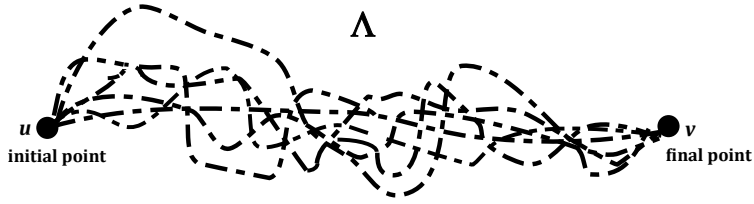

**Fig. 1** Some of the paths that a particle could take in going from **u** to **v**

In this system, based on the FPI, one finds the probability amplitude $\langle v|u\rangle$ for a quantum event[†] whose initial point is *u* and final point is *v* by summing the exponentiated-action, $\exp(iS[\text{path}]/\hbar)$, over each possible path from *u* to *v* [4, 5]. That is, the amplitude is

$$\langle v|u\rangle = \sum_{\text{path } x(w) \text{ from } u \text{ to } v} e^{iS[x(w)]/\hbar} = k\int e^{iS[x(w)]/\hbar}\mathcal{D}x, \qquad (1)$$

where $\mathcal{D}$ is the "integration-measure" on $\Lambda$, and *k* is a normalizing constant independent of the paths (here the action, *S*, is the integral of the difference of kinetic and potential energy over a path, and *h* is Planck's constant).

**(b) The derivations of the Bell inequalities do not make sense over the FPI framework**

(i) So-called "hidden variables" are an underlying, inaccessible mechanism hypothesized to account for the observed features of quantum theory. Bell's theorem purports to show that no local, hidden-variable theory can reproduce the results of quantum mechanics. He writes regarding hidden variables $\lambda$, "It is a matter of indifference in the following [in his proof] whether $\lambda$ denotes a single variable or a set, or even a set of functions, and whether the variables are discrete or continuous [6]."

For a Bell-type experiment in the FPI framework the hidden variables $\lambda$ are the set of all possible paths from the source to the detectors. To see why Bell's argument is meaningless over the FPI system, we begin by quoting the first few steps of his proof [7]:

Suppose that the hypothetical complete description of the initial state is in terms of hidden variables $\lambda$ with probability distribution $\rho(\lambda)$ for the given quantum-mechanical state. The result *A* (= ± 1) of the first measurement can clearly depend on $\lambda$ and on the setting $\alpha$[‡] of the first instrument. Similarly, *B* can depend on $\lambda$ and $\beta$. But our notion of locality requires that *A* does not depend on $\beta$, nor *B* on $\alpha$. We then ask if the mean value … of the product *AB*,

$$E(\alpha,\beta) = \int A(\alpha, \lambda)B(\beta, \lambda)\, \rho(\lambda)d\lambda \qquad (2)$$

can equal the quantum-mechanical prediction.

---

[†] In this paper for simplicity we will only consider events in configuration space, though a similar construct works for spin (see A. Atland, B. Simmons, *Condensed Matter Field Theory*, Cambridge University Press, 2006, Sec. 3.3.5)

[‡] We've replaced Bell's original notation by "$\alpha$," "$\beta$" and "*E*."

Starting with the integral involving $\lambda$ in Eq. 2, Bell considers two further measurements $\alpha'$ and $\beta'$, and quickly derives an inequality that quantum mechanics violates (the complete, simple argument is in Appendix A).

Bell's derivation breaks down in the FPI framework ($\Lambda$, $\mathcal{D}$), however, because "there is no countably-additive measure $\mathcal{D}$ on $C_{u,v}^{0,t}$ [= $\Lambda$] that weighs all paths equally" [and] … the space $C_{u,v}^{0,t}$ is not compact and so its measure would be infinite… [8]." In other words, the FPI framework ($\Lambda$, $\mathcal{D}$) is not a Kolmogorov probability space [9].

**Therefore, the central premise of Bell's argument (the integral $\int A(\alpha, \lambda) B(\beta, \lambda) \, \rho(\lambda) d\lambda$ of Eq. 1) is meaningless in the FPI framework, even though that framework replicates the quantum probabilistic predictions.**

(ii) For similar reasons the CHSH argument [10] also fails to make sense over the FPI framework. The CHSH argument begins by taking the sum and difference of the four products in Eq. 3 below. These are assumed to belong to a local system of physics, where as in (i) $\lambda$ is a presumed set of hidden variables and $A$, $B = \pm 1$ are binary valued functions of the measurement settings $\alpha$, $\beta$, $\alpha'$, $\beta'$. Let $S$ be as follows:

$$S = A(\alpha,\lambda)B(\beta,\lambda) + A(\alpha',\lambda)B(\beta,\lambda) + A(\alpha,\lambda)B(\beta',\lambda) - A(\alpha',\lambda)B(\beta',\lambda). \qquad (3)$$

Using the fact that $A$, $B = \pm 1$, we have that $S = \pm 2$:

$$A(\alpha,\lambda)B(\beta,\lambda) + A(\alpha',\lambda)B(\beta,\lambda) + A(\alpha,\lambda)B(\beta',\lambda) - A(\alpha',\lambda)B(\beta',\lambda)$$
$$= \bigl(A(\alpha,\lambda) + A(\alpha',\lambda)\bigr)B(\beta,\lambda) + \bigl(A(\alpha,\lambda) - A(\alpha',\lambda)\bigr)B(\beta',\lambda) = \pm 2.$$

Taking the integral over each term in Eq. 3, we then have

$$\int_\Lambda A(\alpha,\lambda)B(\beta,\lambda)\rho(\lambda)d\lambda + \int_\Lambda A(\alpha',\lambda)B(\beta,\lambda)\rho(\lambda)d\lambda +$$
$$\int_\Lambda A(\alpha,\lambda)B(\beta',\lambda)\rho(\lambda)d\lambda - \int_\Lambda A(\alpha',\lambda)B(\beta',\lambda)\rho(\lambda)d\lambda$$
$$=$$
$$\int_\Lambda \bigl(A(\alpha,\lambda)B(\beta,\lambda) + A(\alpha',\lambda)B(\beta,\lambda) + A(\alpha,\lambda)B(\beta',\lambda) + A(\alpha',\lambda)B(\beta',\lambda)\bigr)\rho(\lambda)d\lambda \leq 2 \qquad (4)$$

Ineq. 4 is in conflict, however, with the quantum values for certain experiments and measurement values of $\alpha$, $\beta$, $\alpha'$, $\beta'$. In other words, the inequality (which is assumed to be derived over a local system of physics) fails to meet the quantum predictions. This is the standard conclusion.

But, as in (i) above, in order for the derivation of Ineq. 4 to be meaningful, the assumed local system must also meet the conditions of a Kolmogorov probability space. As pointed out above, however, the FPI framework ($\Lambda$, $\mathcal{D}$) is not such a space, though it replicates the quantum probabilistic predictions.

### (c) Brief history of the FPI framework

Of course one might dismiss the system (Λ, $\mathcal{D}$) as representing an inadequate theory because it does not obey the Kolmogorov probability axioms, where one has a sample space Ω, a $\sigma$-algebra $\mathcal{F}$ of subsets of Ω, and a $\sigma$-additive measure $\mu$ on Ω with $\mu(\Omega) = 1$. If one takes that view then Bell's theorem certainly holds, since it's just a derivation of certain elementary inequalities over a well-behaved system admitting a Lebesgue integral [11]. But that would be ignoring more than seventy years of work on the path integral.

Feynman developed the path integral in the early 1940s, when he was still a graduate student at Princeton. His approach quickly led to alternative solutions to various elementary quantum problems that had been solved earlier using Hilbert space and the Schrödinger equation. It took, however, decades longer to arrive at the path-integral description of spin [12] and also at a complete path-integral description of the hydrogen atom (the latter had once symbolized the success of the Schrödinger approach) [13]. Over the years continuing development of the FPI has made it almost indispensable in quantum field theory. But in any case (as stated in Ref. 2, 3) it is now generally recognized that any problem in quantum mechanics that can be solved using Hilbert space and the Schrödinger equation can be solved using just the FPI.

The theoretical justification of the path integral is carried out by time-slicing the corresponding amplitude for a particle traveling from $u$ to $v$, at each stage inserting an infinite resolution of the identity (a so-called *fat identity*) and then taking an integral over all corresponding positions. In the limit (over an infinite product of such integrals) one arrives at the value of the amplitude in Eq. 1, which agrees with the standard Hilbert-space result. The resulting system replicates the quantum probabilistic predictions, though it does not conform to the Kolmogorov axioms. Of course one could reject the validity of the FPI process and choose instead to assume that the setting of Bell's theorem has a nice, well-behaved Lebesgue integral that allows one to derive certain elementary integral-inequalities that quantum mechanics violates. You pays your money you and you takes your choice (from an 1846 Punch cartoon).

On the other hand, if one accepts the FPI formalism the conclusion is that the seeming nonlocalities in quantum mechanics are merely an *artifact of a particular formalism*, as Einstein suspected. But more than that (as we show below), the FPI framework explains—purely in terms of path interference local to each side of a two-particle experiment, with no information sent across the origin—the otherwise mysterious correlations observed for entangled particles. Such an explanation is needed, because the correlations exhibited by such particles do exist experimentally. And, as Bell once noted, "the scientific attitude is that correlations cry out for explanation [14]."

### II. CORRELATIONS VIA A GENETIC HYPOTHESIS

So far we have shown that Bell-type arguments break down in the FPI system (Λ, $\mathcal{D}$), although the FPI system replicates the quantum-mechanical predictions. In this section we show that regardless of whether a system is classical or quantum, when there is a genetic hypothesis no information needs to be sent across the origin to produce the correlations.

Here we have borrowed the term "genetic" from Bell. As he once told Jeremy Bernstein [15],

The discomfort that I feel is associated with the fact that the observed perfect quantum correlations seem to demand something like the "genetic" hypothesis. For me, it is so reasonable to assume that the photons in those experiments carry with them programs, which have been correlated in advance, telling them how to behave. This is so rational that I think that when Einstein saw that, and the others refused to see it, *he* was the rational man. The other people, although history has justified them, were burying their heads in the sand. I feel that Einstein's intellectual superiority over Bohr, in this instance, was enormous; a vast gulf between the man who saw clearly what was needed, and the obscurantist. So for me, it is a pity that Einstein's idea doesn't work. The reasonable thing just doesn't work.

But the reasonable thing does work after all in ($\Lambda$, $\mathcal{D}$), as we show below.

For example, consider David Mermin's famous red-green version of Bell's Theorem [16]. In Mermin's setup, two particles travel out in opposite directions from a source to where "measurements" are then carried out on each particle before it reaches the detectors. There are three possible measurement settings, each setting causing a detector to flash red or green. Both particles at the source event carry identical "instruction sets" consisting of triples of red and green that deterministically govern whether a red or green light flashes at a detector. The probability of agreement between the sides produced by the instruction sets is at least 5/9, as Mermin shows. But he then writes,

> Therefore if instruction sets exist, the same colors will flash in at least 5/9 of all the runs, regardless of how the instruction sets are distributed from one run of the demonstration to the next. This is Bell's theorem (also known as Bell's inequality) for the *gedanken* demonstration. ... But in the actual *gedanken* demonstration [quantum experiment] the same colors flash only ½ the time. [This violates] Bell's inequality, and therefore there can be no instruction sets.

Well, not quite. True, the red-green instruction sets do not produce a result identical to that of quantum mechanics, but they do produce correlations between the sides without any information being sent across the origin. They constitute a valid "genetic hypothesis."

We can bring about a similar result in a thought experiment that is a limited version of the one that we describe for ($\Lambda$, $\mathcal{D}$) below. Thus suppose that each particle travels out from the source along a single path—the path of least action—and associate with it a clock $\exp(iS[path]$§$)$ that rotates along with it uniformly to where a measurement is carried out. Let a measurement consist of altering the path length so that the clock "ticks" further by 0, $2\pi/3$, or $4\pi/3$ radians from what it otherwise would. Assume the clocks associated with the pair of particles are synchronized at the source event, and suppose that on each side a detector $A(\theta)$ signals +1 if the "clock" angle $\theta$ (i.e. $S[path]$) of the arriving particle is, say, between 0 and $\pi$, and –1 if it is between $\pi$ and $2\pi$. It's easy to prove that this genetic hypothesis (a clock associated with each of the two particles) produces, like Mermin's instruction sets, a higher probability of agreement (⅔) when the settings are different between the sides than does a similar quantum experiment.

---

§ We can let $\hbar$ = 1 here

Of course the quantum result in an analogous setup is smaller, but the point here is that the identical, synchronized "clock" instructions moving uniformly along each of the two paths—the genetic hypothesis—produce correlations in such a way that no information needs to be sent across the origin.

To see how the genetic hypothesis works in a similar quantum setup in configuration space, we now look at a two-particle Rarity-Tapster interferometer [17], where a mother particle down-converts (in a gedanken fashion) to two non-relativistic, entangled daughter particles. The figure shows a few of the infinitely many paths (the "hidden variables") on each side from the source to the detectors. An opaque barrier separates the two sides at the source, so that all path interference is local to the respective sides. In such an experimental arrangement the source cannot be a point source, as it can be in a single-particle interferometer [18]. There is always positional and temporal uncertainty at the source, which implies that there are always more than two possible paths from the source to the beam splitters, though conventional pictures show only two such paths.

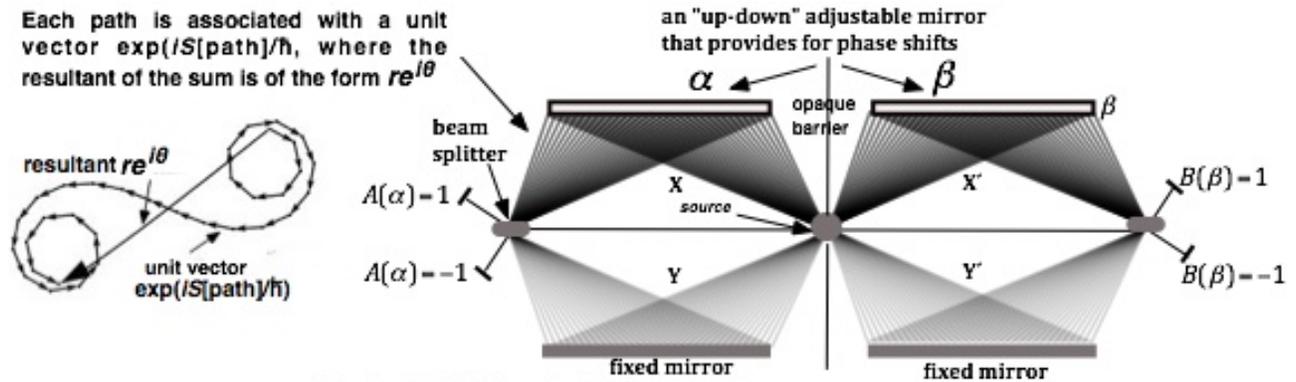

Fig. 2 Rarity-Tapster interferometer

The genetic hypothesis in this case consists of clock instructions—unit vectors $\exp(iS[path]/\hbar)$—associated with each of the paths, the clocks rotating the same over congruent paths. This is similar to the genetic hypothesis above based on the clock instructions for the case of a single path on a side. In the single-path setup the direction of the clock-angle $\theta$ in the exponentiated-action (when the particle arrives at a detection point) determines the values of $A(\theta)$, $B(\theta) = \pm 1$. This explains the correlations without any information being sent across the source.

In the quantum case we can explain the correlations similarly (no information needing to be sent across the origin) by considering the resultant $re^{i\theta}$ of the sum of unit vectors $\exp(iS[path]/\hbar)$ associated with all paths on each side. We postulate that the resultant's angle, $\theta$, determines (in $A(\theta)$ and $B(\theta)$ as before) whether transmittal or reflection occurs at a beam splitter. In the standard Hilbert-space setup a global phase factor is considered physically meaningless since multiplying by such a constant leaves the result statistically invariant. But in the FPI framework we can clearly treat the resultant $re^{i\theta}$ as a fixed outcome when the various unit vectors are summed. The resultant is a consequence of Eq. 1 in Sec. I. This postulate is obviously consistent with experiment, and not only can it not be violated by experiment, it provides a genetic hypothesis that explains the correlations. Be that as it may, the outcomes occur via path-interference local to each side. Moreover, they are in agreement with the standard Hilbert-space predictions.

Thus we have shown that the supposed nonlocalities in quantum mechanics are merely an artifact of the particular formalism used to represent quantum phenomena: they do not exist in the FPI framework, which faithfully replicates the quantum predictions. **QED**

## APPENDIX A: Bell's Proof

Starting from Eq 1 in Sec. I, Bell lets $\alpha'$, $\beta'$ be alternate settings for the measurements while using $|A|$, $|B| \leq 1$ and $\int \rho(\lambda)d\lambda = 1$ (the steps are on p. 37 of [19]). Thus

$$E(\alpha,\beta) - E(\alpha,\beta') = \int [A(\alpha,\lambda)B(\beta,\lambda) - A(\alpha,\lambda)B(\beta',\lambda)]\rho(\lambda)d\lambda$$
$$= \int [A(\alpha,\lambda)B(\beta,\lambda)[1 \pm A(\alpha',\lambda)B(\beta',\lambda)]\rho(\lambda)d\lambda - \int [A(\alpha,\lambda)B(\beta',\lambda)[1 \pm A(\alpha',\lambda)B(\beta,\lambda)]\rho(\lambda)d\lambda$$

Thus, by the triangle inequality and the fact that

$$0 \leq 1 \pm A(\alpha',\lambda)B(\beta',\lambda),\ 0 \leq 1 \pm A(\alpha',\lambda)B(\beta,\lambda),\ 1 = \int 1\rho(\lambda)d\lambda$$

we have

$$\left| E(\alpha,\beta) - E(\alpha,\beta') \right| \leq \left| \int [1 \pm A(\alpha',\lambda)B(\beta',\lambda)]\rho(\lambda)d\lambda + \int [1 \pm A(\alpha',\lambda)B(\beta,\lambda)]\rho(\lambda)d\lambda \right|$$
$$\leq 2 \pm \left[ \int [A(\alpha',\lambda)B(\beta',\lambda)]\rho(\lambda)d\lambda + \int [A(\alpha',\lambda)B(\beta',\lambda)]\rho(\lambda)d\lambda \right]$$
$$= 2 \pm \left[ E(\alpha',\beta') + E(\alpha',\beta) \right],$$

which is violated by the quantum predictions for certain settings of $\alpha$, $\beta$ $\alpha'$, $\beta'$.

---

[1] D. Albert, R. Galchen, *Was Einstein Wrong? A Quantum Threat to Special Relativity,* Scientific American Magazine, March 2009

[2] M. S. Swanson. *Path Integrals and Quantum Processes*, §2.2

[3] S. Weinberg, *Lectures on Quantum Mechanics*, Ch. 9, Cambridge University Press, 2012

[4] R. P. Feynman, A. R. Hibbs, *Quantum Mechanics and Path Integrals*, emended by D. Styer

[5] T. Lancaster, S. J. Blundell, *Quantum Field Theory for the Gifted Amateur*, Oxford University Press

[6] J. S. Bell, "On the Einstein-Podolsky-Rosen paradox," in *Speakable and Unspeakable in Quantum Mechanics*

[7] J. S. Bell, "Introduction to the hidden-variable question," in *Speakable and Unspeakable in Quantum Mechanics*

[8] G. Johnson and M. Lapidus, *The Feynman Integral and Feynman's Operational Calculus*, pp. 32-33, 102-103. The proofs are quite elementary.

[9] L. Koralov, Y. Sinai, *Theory of Probability and Random Processes*, Ch 1, Springer

[10] J. F. Clauser, M. A. Horne, A. Shimony, R. Holt, October 1969, PhysRevLett. 23

[11] A. Khrennikov has a rigorous proof of Bell's theorem over a Kolmogorov space in *A mathematicians viewpoint to Bell's theorem: In memory of Walter Phillip*